\documentclass[letter,traditabstract]{aa} 		

\usepackage{graphicx}
\usepackage{txfonts}
\usepackage{natbib}
\usepackage{enumitem}

\begin{document}

\title{Dissecting the cosmic infra-red background with Herschel/PEP
\thanks{Herschel is an ESA space observatory with science instruments provided by 
European-led Principal Investigator consortia and with important participation from NASA.}}

\author{S. Berta\inst{1}
        \and
        B. Magnelli\inst{1}
	\and
	D. Lutz\inst{1}
	\and 
	B. Altieri\inst{2}
	\and
	H. Aussel\inst{3}
	\and
	P. Andreani\inst{4}\fnmsep\inst{5}
	\and
	O. Bauer\inst{1}
	\and
	A. Bongiovanni\inst{6}\fnmsep\inst{7}
	\and
	A. Cava\inst{6}\fnmsep\inst{7}
	\and
	J. Cepa\inst{6}\fnmsep\inst{7}
	\and
	A. Cimatti\inst{8}
	\and
	E. Daddi\inst{3}
	\and
	H. Dominguez\inst{9}
	\and
	D. Elbaz\inst{3}
	\and 
	H. Feuchtgruber\inst{1}
	\and
	N.M. F{\"o}rster Schreiber\inst{1}
	\and
	R. Genzel\inst{1}
	\and
	C. Gruppioni\inst{9}
	\and
	R. Katterloher\inst{1}
	\and
	G. Magdis\inst{3}
	\and
	R. Maiolino\inst{10}
	\and 
	R. Nordon\inst{1}
	\and
	A.M. P{\'e}rez Garc{\'\i}a\inst{6}\fnmsep\inst{7}
	\and 
	A. Poglitsch\inst{1}
	\and
	P. Popesso\inst{1}
	\and 
	F. Pozzi\inst{8}
	\and
	L. Riguccini\inst{3}
	\and
	G. Rodighiero\inst{11}
	\and 
	A. Saintonge\inst{1}
	\and
	P. Santini\inst{10}
	\and
	M. Sanchez-Portal\inst{2}
	\and
	L. Shao\inst{1}
	\and
	E. Sturm\inst{1}
	\and
	L.J. Tacconi\inst{1}
	\and 
	I. Valtchanov\inst{2}
	\and 
	M. Wetzstein\inst{1}
	\and
	E. Wieprecht\inst{1}
        }

\offprints{Stefano Berta, \email{berta@mpe.mpg.de}}

\institute{\centering \vskip -10pt \small \it (See online Appendix \ref{sect:affiliations} for author affiliations) }

   \date{Received 31 March 2010; accepted 21 April 2010}

 
  \abstract{
  The constituents of the cosmic IR background (CIB) are studied at 
  its peak wavelengths (100 and 160 $\mu$m) by exploiting Herschel/PACS observations of the GOODS-N, Lockman Hole, and COSMOS
  fields in the {\em PACS Evolutionary Probe} (PEP) guaranteed-time survey.
  The GOODS-N data reach 3$\sigma$ depths of 
  $\sim$3.0 mJy at 100 $\mu$m and $\sim$5.7 mJy at 160 $\mu$m. At these levels, 
  source densities are 40 and 18 beams/source, respectively, thus hitting the confusion limit at 160$\mu$m.
  Differential number counts extend from a few mJy up to 100-200 mJy, and are approximated as a double 
  power law, with the break lying between 5 and 10 mJy.
  The available ancillary information allows us to split 
  number counts into redshift bins. At $z\le0.5$ we isolate a class of luminous sources  
  ($L_{IR}\sim10^{11}$ L$_\odot$), whose SEDs resemble late-spiral galaxies, peaking at $\sim$130 
  $\mu$m restframe and significantly colder than what is expected on the basis of pre-Herschel models.
  By integrating number counts over the whole covered flux range, we obtain a surface brightness of 
  $6.36\pm1.67$ and $6.58\pm1.62$ $[$nW m$^{-2}$ sr$^{-1}]$ at 100 and 160 $\mu$m, resolving  
  $\sim$45\% and $\sim$52\% of the CIB, respectively. 
  When stacking 24 $\mu$m sources, the inferred CIB lies within 1.1$\sigma$ and 0.5$\sigma$ from 
  direct measurements in the two bands, and fractions increase to 50\% and 75\%.
  Most of this resolved CIB fraction was radiated at $z\le1.0$,
  with 160 $\mu$m sources found at higher redshift than 100 $\mu$m ones.
  }


   \maketitle


\section{Introduction}

The cosmic IR background \citep[CIB,][]{puget1996,hauser1998} accounts 
for roughly half of the total 
extragalactic background light \citep[EBL,][]{hauser2001,lagache2005}, 
i.e., half of the energy radiated by all galaxies, at all cosmic epochs, at 
any wavelength \citep{dole2006}.
It is therefore a crucial constraint on modes and times of galaxy formation.

Deep cosmological surveys carried out with the {\em Infrared Space Observatory} (ISO, see 
Genzel \& Cesarsky \citeyear{genzel2000} for a summary) and {\em Spitzer Space Telescope} 
\citep[][for a review]{soifer2008}
produced large samples of mid-IR sources and deep number counts \citep{elbaz2002,papovich2004}. 
These surveys led to mid-IR CIB lower limits within a factor of two from the upper constraints 
set by TeV cosmic opacity measurements \citep[e.g.][]{franceschini2009}.

However, at CIB peak wavelengths ($100-200$ $\mu$m), the nature of individual galaxies 
building up the EBL is poorly known. Past surveys produced limited samples of distant far-IR
objects \citep[e.g.][]{frayer2009}, mainly due to the small apertures of the available 
instruments and the low sensitivity in the far-IR. 
In the 160 $\mu$m Spitzer/MIPS band, $\sim$7\% of the CIB was resolved into individually detected objects 
\citep{dole2004}, and it was only through stacking the 24 $\mu$m sources that 
most ($60-70$\%) of the far-IR CIB could be recovered \citep{dole2006,bethermin2010}.
Similarly, at longer wavelengths, only stacking of 24 $\mu$m sources on BLAST maps 
could account for the majority of the EBL at 250, 350, and 500 $\mu$m \citep{marsden2009}.

With the favorable diffraction limit of the large {\em Herschel} 3.5 m mirror (Pilbratt et al., \citeyear{pilbratt2010}), 
and the high sensitivity of its {\em Photodetector Array Camera \& Spectrometer} (PACS, 
70, 100, 160 $\mu$m; Poglitsch et al., \citeyear{poglitsch2010}), confusion and blending of sources are much less of a limitation.
We are now able to resolve a large fraction of the CIB at its peak into individual galaxies.

The {\em PACS Evolutionary Probe} (PEP) 
extragalactic survey samples 4 different tiers:
from the wide and shallow COSMOS field, through medium size areas like the Lockman Hole, 
all the way down to the 160 and 100 $\mu$m confusion limit in the pencil-beam, 
very deep observations in GOODS-N and GOODS-S, and even beyond by exploiting gravitational lensing 
in low-redshift galaxy clusters (e.g. Abell 2218, Altieri et al. \citeyear{altieri2010}).

\begin{figure*}[!ht]
\centering
\rotatebox{-90}{
\includegraphics[height=0.425\textwidth]{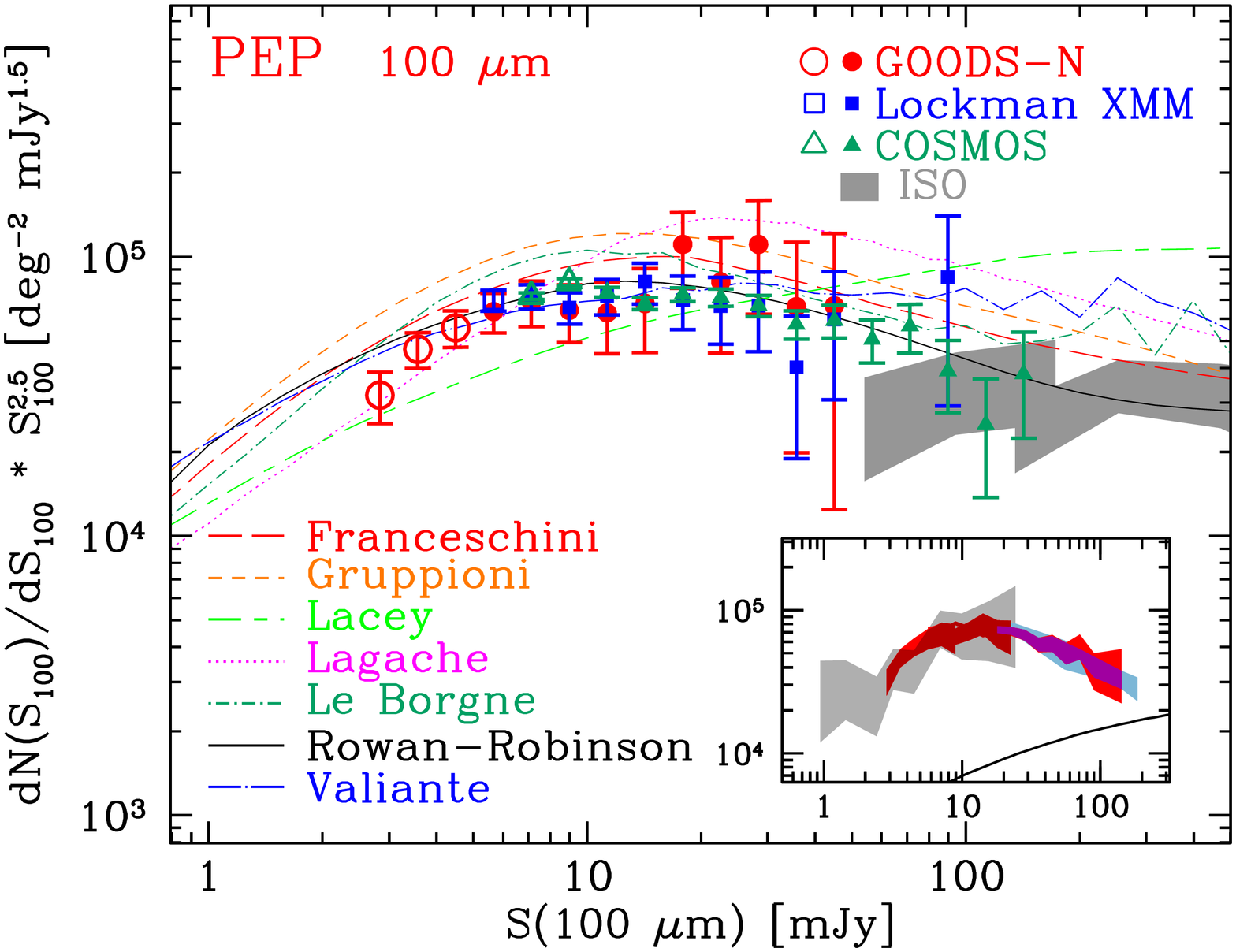}
}
\rotatebox{-90}{
\includegraphics[height=0.425\textwidth]{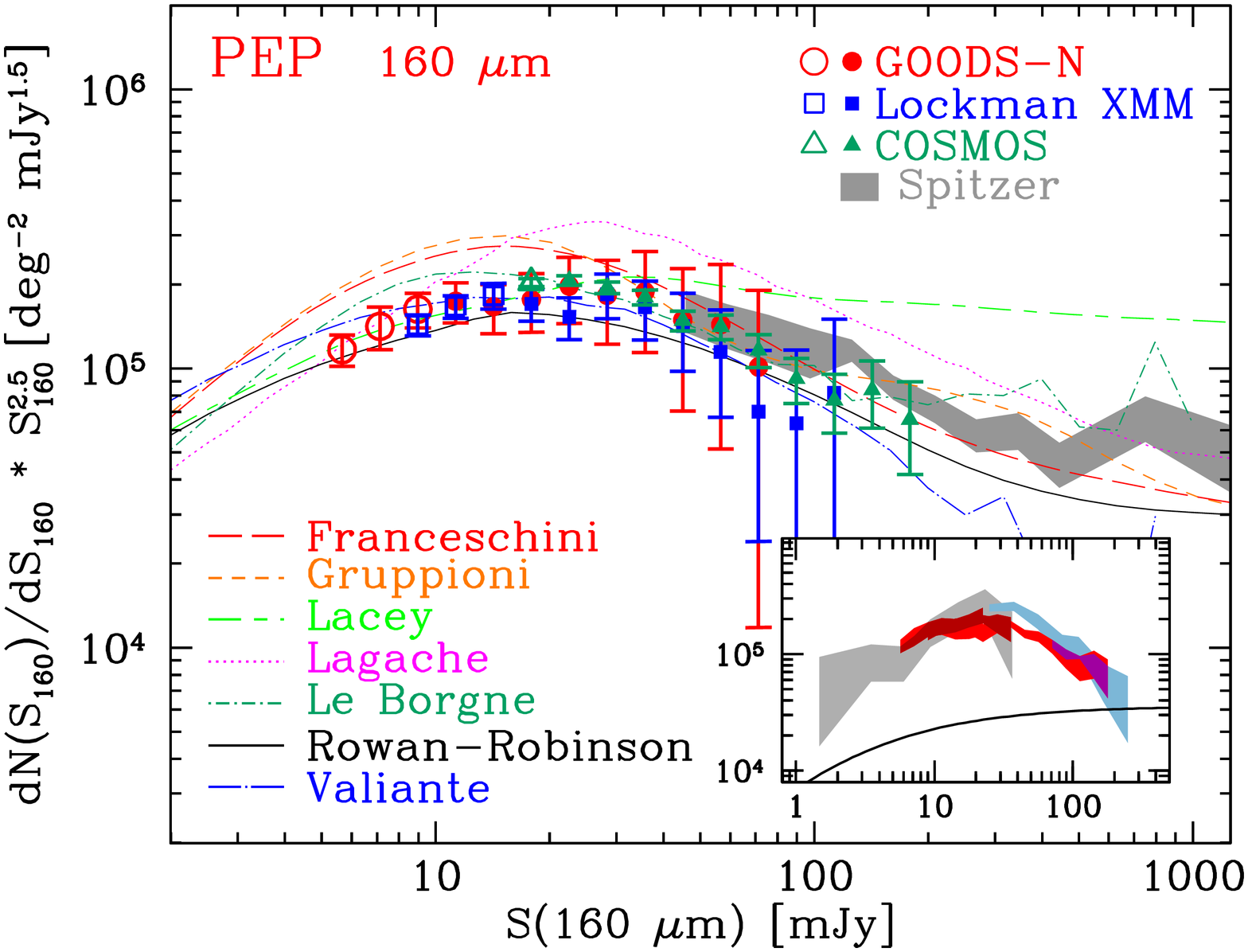}
}
\caption{Number counts at 100 ({\em left}) and 160 $\mu$m ({\em right}), 
normalized to the Euclidean slope. Filled/open symbols belong to flux bins above/below the 80\% 
completeness limit. Models belong to
\citet{lagache2004}, \citet{franceschini2009}, \citet{rowanrobinson2009}, 
\citet{leborgne2009}, \citet{valiante2009}, \citet{lacey2009}, Gruppioni \& Pozzi (in prep.).
Shaded areas represent ISO and Spitzer data \citep{rodighiero2004,heraudeau2004,bethermin2010}.
{\em Inset}: Collection of PACS number counts, including this work (red, limited to relative errors 
$\le$15\% for clarity sake), PEP Abell 2218 (grey, 
Altieri et al. \citeyear{altieri2010}) and HerMES-PACS (light-blue, Aussel et al. \citeyear{aussel2010}). 
The solid lines in the insets mark the trends expected for 
a non-evolving population of galaxies.}
\label{fig:pep_counts}
\end{figure*}

Here we exploit the {\em science demonstration phase} (SDP) observations
of the GOODS-N field, complemented with the COSMOS 
and Lockman Hole wider and shallower layers of the survey, to build far-IR galaxy number 
counts and derive the fraction of CIB resolved by PEP.
We then take advantage of the extensive multi-wavelength coverage of the GOODS-N
field, with the aim of identifying the CIB contributors and the epoch when the bulk of 
the Universe IR energy budget was emitted.


\section{Number counts}\label{sect:counts}

The area used to derive number counts includes observations in the GOODS-N ($\sim$140 arcmin$^2$),
COSMOS ($\sim$2 deg$^2$) and Lockman Hole (LH, $\sim$450 arcmin$^2$) fields. 
Appendix \ref{sect:data} describes data reduction and catalog extraction. 

PACS number counts were computed in each field 
within 0.1 dex flux bins, and are reported in  
Fig. \ref{fig:pep_counts} and Table \ref{tab:pep_counts}, normalized to the Euclidean 
slope ($dN/dS\propto S^{-2.5}$).
Incompleteness and spurious sources have been dealt with 
using Monte Carlo simulations (see Appendix \ref{sect:data}). Filled symbols 
belong to the conservative 80\% completeness limit. Thanks to the method described in \citet{chary2004},
we extend the analysis down to the faintest objects detected at the 3$\sigma$ level. 
Error bars include 
Poisson statistics, flux calibration uncertainties, and photometric errors.
The latter have been 
propagated into number counts via $10^4$ 
realizations of random Gaussian flux errors applied to each PACS source, 
using a dispersion equal to the local measured noise.
The 95 $\mu$m ISO counts \citep{rodighiero2004,heraudeau2004} and 
the 160 $\mu$m Spitzer results
by \citet[][including GOODS/FIDEL, COSMOS and SWIRE fields]{bethermin2010}
are consistent with PACS data, excepted the ISO faintest flux bin.

Small insets in Fig. \ref{fig:pep_counts} show results from PACS Guaranteed-Time surveys:
this work, Abell 2218 lensed counts (Altieri et al. \citeyear{altieri2010}), and 
HerMES (Aussel et al. \citeyear{aussel2010}).
The three datasets complement each other and show overall agreement, the 160 $\mu$m HerMES counts excepted. 
This discrepancy can be ascribed to the sheet of galaxies known to bias the HerMES 
LH {\em North} field, on which their counts are based \citep{owen2009}.

Differential, Euclidean-normalized counts display various well-known features: below 
$\sim$100 mJy at 100 $\mu$m and $\sim$200 mJy at 160 $\mu$m, 
the slope is super-Euclidean, reaching a peak in the counts between $\sim 5-10$ and 30 mJy, 
while at fainter fluxes the slope quickly becomes sub-Euclidean. These attributes
are generally interpreted as indicating evolving properties in  
the IR galaxy population (see Sect. \ref{sect:discussion}): 
the trend expected for no-evolution is shown in Fig. \ref{fig:pep_counts} insets.

The non-normalized differential counts can be represented by a power law of the type
$dN/dS\propto S^{\alpha}$, with two distinct slopes at fluxes fainter/brighter 
than $S_{break}$. The results of a weighted least squares fit are given in Table 
\ref{tab:pow_law} for the different PEP fields. The $S_{break}$ flux
is $\sim$5.0 mJy at 100 $\mu$m and 
$\sim$8.5 mJy at 160 $\mu$m. At the bright end, the number of sources in the small fields is low, 
and error bars are dominated by Poisson statistics. 
Consequently, the large uncertainties allow for nearly-Euclidean fits in 
GOODS-N and the Lockman Hole. In the case of COSMOS, the slope is significantly 
steeper than Euclidean at 160 $\mu$m, while the 100 $\mu$m counts are still 
very flat. 
At the faint end, the 160 $\mu$m slope derived from our counts (GOODS-N only) is consistent with what is 
found by stacking 24 $\mu$m sources on Spitzer 160 $\mu$m maps \citep[$\alpha=-1.61\pm0.21$]{bethermin2010}.

\setcounter{table}{1}

\begin{table}[!ht]
\centering
\begin{tabular}{l c c c}
\hline
\hline
Field & Flux range  & Slope & Error \\
band  & $[$mJy$]$ & $\alpha$ & $d\alpha$\\
\hline
GOODS-N 100 & 2.8$-$5.6 & -1.32 & $\pm$0.27\\
GOODS-N 100 & 5.6$-$45 & -2.31 & $\pm$0.05\\
LH 100 & 6.0$-$89 & -2.60 & $\pm$0.05\\
COSMOS 100 & 8.0$-$142 & -2.63 & $\pm$0.03\\
\hline
GOODS-N 160 & 5.6$-$9.0 & -1.67 & $\pm$0.32\\
GOODS-N 160 & 9.0$-$71 & -2.54 & $\pm$0.04\\
LH 160 & 9.0$-$113 & -2.68 & $\pm$0.05\\
COSMOS 160 & 17.0$-$179 & -3.02 & $\pm$0.04\\
\hline
\end{tabular}
\caption{Power-law fit to PACS differential number counts in the form $dN/dS\propto S^{\alpha}$.
}
\label{tab:pow_law}
\end{table}


\section{Resolved CIB fraction}\label{sect:cib}

The added contribution of resolved galaxies provides a lower limit to the IR 
background and can be compared to direct measurements of the total CIB. 
After the discovery of the CIB \citep{puget1996,hauser1998}, numerous authors attempted 
to directly measure 
its surface brightness from COBE/DIRBE maps 
\citep[e.g.][]{lagache2000,renault2001,wright2004,dole2006}. Here we adopt the most recent 
revision of the \citet{lagache2000} DIRBE measurements, provided in \citet{dole2006}: 
$14.4\pm6.3$, $12.0\pm6.9$, and $12.3\pm2.5$ $[$nW m$^{-2}$ sr$^{-1}]$, at 100, 140, and 240 $\mu$m, respectively.
Interpolating between these values, one obtains a value of $12.8\pm6.4$ $[$nW m$^{-2}$ sr$^{-1}]$ at 160 $\mu$m.
These direct measurements are still affected by large uncertainties, mainly due to difficulties in defining
an absolute flux scale and in removing zodiacal light. An alternative estimate of the total 
CIB is obtained by integrating the power-law extrapolation of our number counts, between 0.01 and 1000
mJy. We adopt the slope derived on GOODS-N data for $S<S_{break}$ and an average between COSMOS and 
Lockman Hole at brighter fluxes (see Table \ref{tab:pow_law}
and Sect. \ref{sect:counts}). The total CIB estimates from this extrapolation are
$\sim$12 and $\sim$13 $[$nW m$^{-2}$ sr$^{-1}]$
in the green (100 $\mu$m) and red (160 $\mu$m) bands, within the errors of the
direct measurements. 

\begin{figure}[!ht]
\centering
\includegraphics[width=0.45\textwidth]{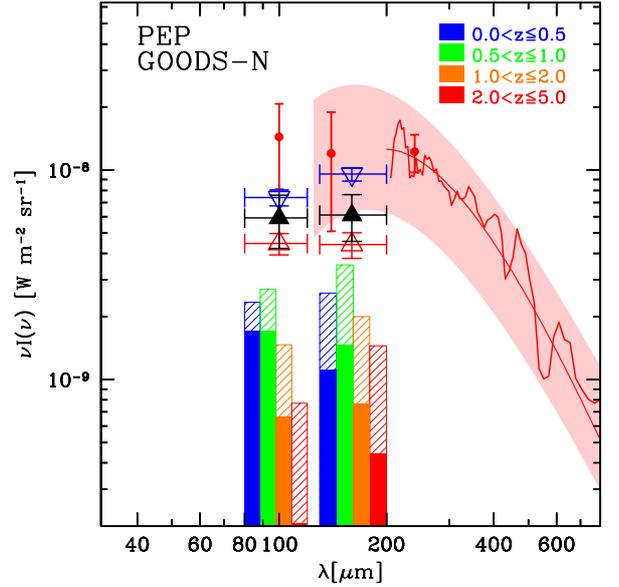}
\caption{Contribution of GOODS-N galaxies to the CIB. 
Red empty triangles represent the sum of individually detected sources; black filled triangles 
come from number counts above the 3$\sigma$ limit, and include completeness correction; 
blue upside-down empty triangles belong to stacking of 24 $\mu$m sources. Solid histograms denote 
the contribution of different redshift bins to the CIB, accounting for
completeness correction. Hatched histograms are stacking results.
Literature data include: DIRBE measurements 
revised by \citet[][filled circles, 1$\sigma$ errors]{dole2006}; FIRAS spectrum 
\citep[solid lines above 200 $\mu$m,][]{lagache1999,lagache2000};  
\citet{fixsen1998} modified Black Body (shaded area).
}
\label{fig:pep_cib}
\end{figure}

The contribution to the CIB of individual GOODS-N sources detected above 3$\sigma$ 
is $\nu I_\nu = 4.46\pm0.52$ at 100 $\mu$m
and $4.41\pm 0.62$ $[$nW m$^{-2}$ sr$^{-1}]$ at 160 $\mu$m, i.e., 
$\sim31\pm4$\% and $\sim35\pm5$\% of the direct DIRBE measurements (Fig. \ref{fig:pep_cib}).
Taking the effects of completeness into account (see Sect. \ref{sect:counts}), 
we can compute the CIB fraction due to the general FIR-population, 
overriding source extraction losses. We limit the analysis to the 
3$\sigma$ detection limit again. The GOODS-N number counts 
cover the ranges $3-45$ mJy (at 100 $\mu$m) and $5.5-72$  mJy (at 160 $\mu$m); 
the COSMOS field allows upper boundaries to be extended to 142 and 179 mJy. 
Integrating the counts over the whole flux range, we obtain $\nu I_\nu=6.36\pm1.67$ and $6.58\pm1.62$ 
$[$nW m$^{-2}$ sr$^{-1}]$, within 1.3$\sigma$ and 1.0$\sigma$ from the reference values at 100 and 160 
$\mu$m, respectively (Fig. \ref{fig:pep_cib}).
These correspond to $\sim 45\pm12$\% and $\sim52\pm13$\% of the \citet{dole2006} 
estimate. 
The derived uncertainty now also includes 
the effect of Poisson statistics and not only photometric errors.
The bright end of number counts, covered only by COSMOS, gives a small 
contribution ($\sim4-6$\%) to the total CIB surface brightness.

The PACS detection rate of Spitzer 24 $\mu$m sources \citep[$S_{24}\ge20$ $\mu$Jy,][]{magnelli2009} 
is roughly 15\%. It is possible to 
derive a deeper CIB estimate, through stacking 
on the PEP maps at the positions of all 24 $\mu$m objects, including those not detected 
in the FIR \citep[e.g.][]{dole2006}. Uncertainties on the stacked 
fluxes are computed via a simple bootstrap procedure.
The CIB surface brightness produced by 24 $\mu$m sources is 
$7.39\pm0.48$ and $9.57\pm 0.71$ $[$nW m$^{-2}$ sr$^{-1}]$
at 100 and 160 $\mu$m, consistent with \citet{bethermin2010}, 
and providing $\sim$51\% and $\sim$75\% of the total background, within 
1.1$\sigma$ and 0.5$\sigma$ from \citet{dole2006}.

\section{Discussion}\label{sect:discussion}

\begin{figure*}[!ht]
\centering
\includegraphics[height=0.4\textwidth]{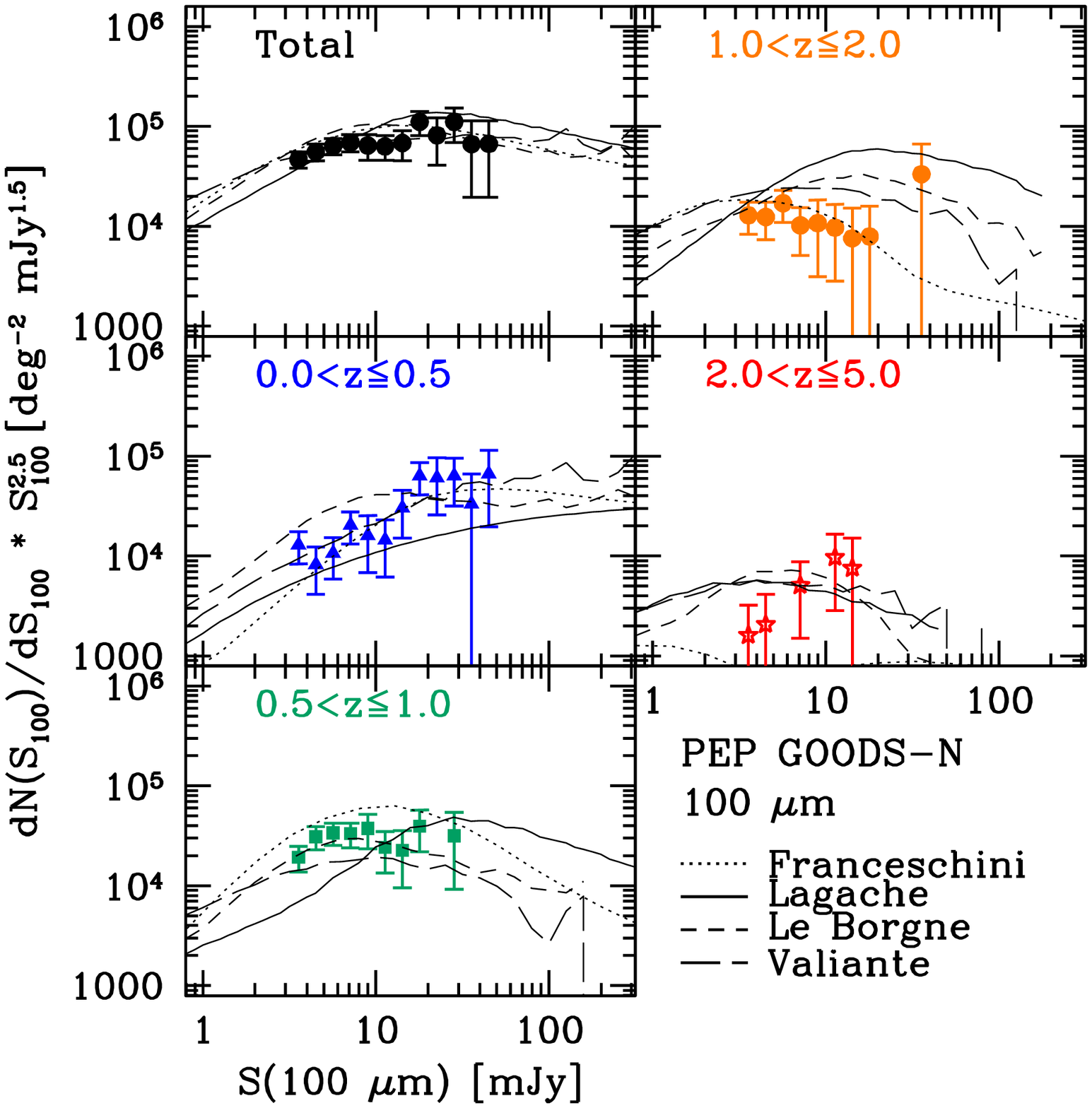}
\includegraphics[height=0.4\textwidth]{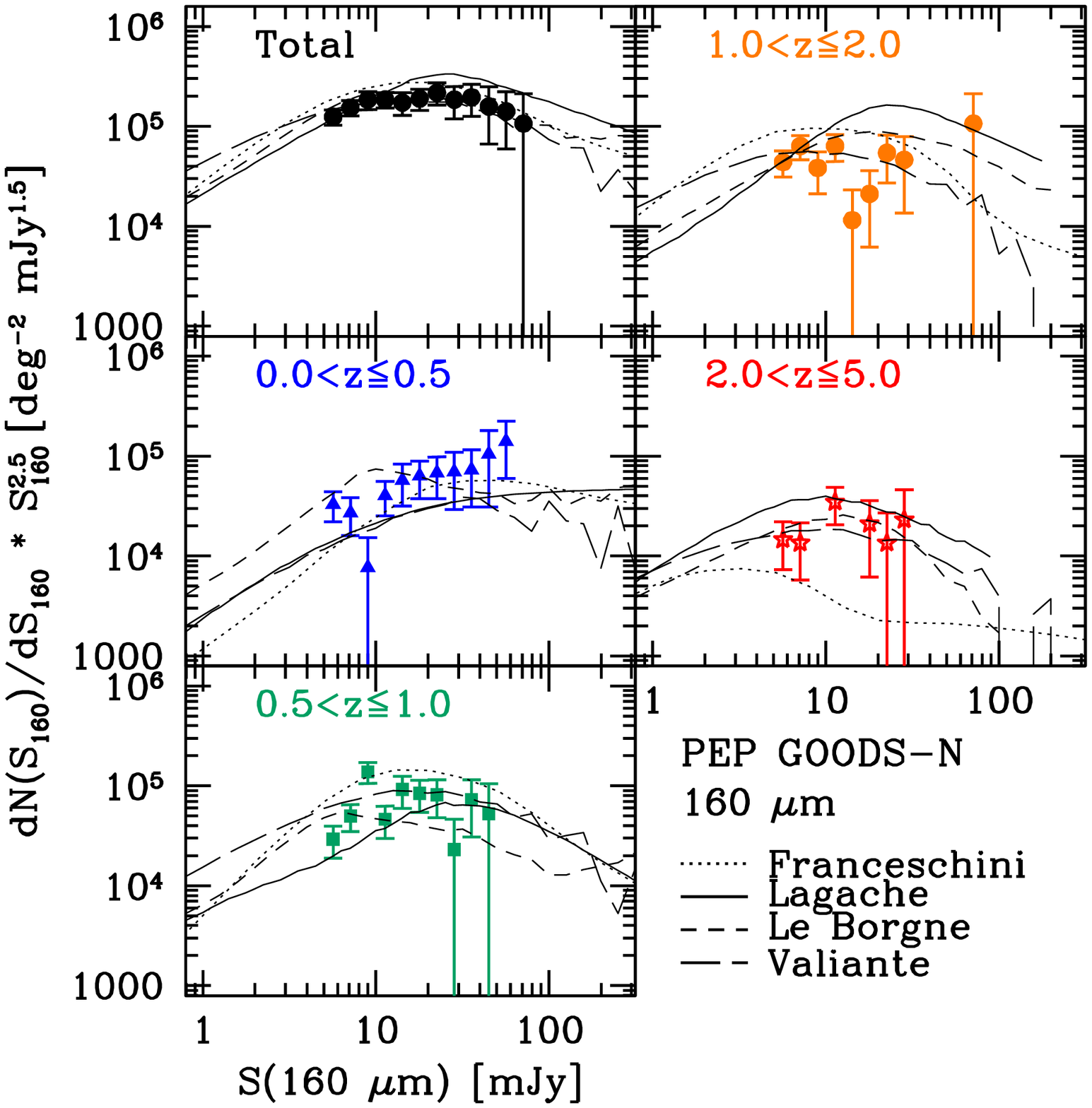}
\caption{Slicing the PEP GOODS-N population into redshift bins. Each PACS source
has been associated to an optical/mid-IR counterpart as described in Appendix \ref{sect:data}.
See Fig. \ref{fig:pep_counts} and text for details on the models shown.}
\label{fig:pep_counts_redshift}
\end{figure*}

In the attempt to reproduce the observed ISO and Spitzer number counts, 
several authors built ``backward'' evolutionary models, including 
luminosity and/or density evolution, as well as different galaxy populations.  
In Fig. \ref{fig:pep_counts}, we overlay recent models 
onto the observed PACS counts. We include in this collection also 
the \citet{lacey2009} $\Lambda$-CDM semi-analytical model (SAM), 
complemented with radiative transfer dust reprocessing. 

The most successful models are the \citet{valiante2009}, including 
luminosity-dependent distribution functions for the galaxy IR SEDs 
and their AGN contribution, and \citet{rowanrobinson2009}, employing 
analytic evolutionary functions without discontinuities and 4 galaxy populations.

Gruppioni and \citet{lagache2004} overestimate 
the amplitude of the number counts peak in both bands, while 
\citet{franceschini2009} and \citet{leborgne2009} reproduce 
the counts fairly well only in one channel (100 and 160 $\mu$m, respectively).
It is worth recalling that to date most of these 
models have been fine-tuned to reproduce mid-IR and sub-mm statistics, while a 
big gap in wavelength was affecting far-IR predictions. Most include a luminosity evolution 
$\propto(1+z)^{3.0-3.5}$, but the redshift limit for this slope, the details of 
density evolution, or the adopted galaxy zoo vary significantly from author to author.

Besides spanning a much wider range of 
observational data (UV, optical, near-IR luminosity functions, galaxy 
sizes, metallicity, etc.), the SAM approach suffers here for a limited flexibility
in the choice of parameters, and significantly overestimates the bright end of 
PACS counts. Moreover, it cannot reproduce the peak, especially in the green band.

Thanks to the rich ancillary dataset in GOODS-N, we split the far-IR number counts 
into redshift bins (Fig. \ref{fig:pep_counts_redshift}).
This elaboration offers a remarkable chance to set 
detailed constraints on the evolution of the 
galaxy populations adopted in current recipes.
This view highlights some new features and the main problems of the models 
under discussion. 

First, in the lowest redshift 
bin, $0.0<z\le0.5$, the differential number counts, normalized to the Euclidean slope,
monotonically increase as a function of flux, resembling the trend 
expected for a non-evolving population of galaxies. 
The consistency of models with data weakens at longer wavelengths.
At the bright end, above 10 mJy and 20 mJy at 160$\mu$m, PEP catalogs include
43 and 18 objects in this redshift range.
Based on template fitting (Rodighiero et al., \citeyear{rodighiero2010}),
half of this sample shows a late-spiral SED ({\em Sd-Sdm} in the Polletta et al., \citeyear{polletta2007}, 
template library) with typical luminosities above $10^{11}$ L$_\odot$ and a peak wavelength of 
$\sim$130 $\mu$m.
Nevertheless, at these luminosities, \citet{chary2001} models, 
which adopt the local $L-T$ relation, predict SEDs peaking around 90 $\mu$m restframe, 
thus indicating that this class is indeed colder than typical local IR-luminous sources.
This is reminiscent of the cold sources found by ISO FIRBACK and serendipity 
surveys \citep[e.g.][]{dennefeld2005,stickel2007}.

Significant differences between models arise at higher redshift.
The bin $1.0<z\le2.0$ shows the largest 
discrepancies: apparently, most of the overestimation 
of total number counts by some models originates here, suggesting that 
the adopted evolution is either too protracted in time or too steep.

We are now able for the first time to infer how much of the resolved CIB peak 
was emitted at different cosmic epochs. Figure 
\ref{fig:pep_cib} depicts the amount of extragalactic IR background originating 
from the four redshift bins taken into account (see also Fig. \ref{fig:pep_counts_redshift}).
At 100 $\mu$m, at the GOODS-N 3$\sigma$ limit of $\sim$3 mJy, roughly 80\% of the 
resolved CIB is emitted by objects detected at $z\le1$, equally distributed below and above 
$z=0.5$. The remainder belongs to more distant 
galaxies, with only 5\% locked into individual objects at $z>2$. 
Most of the resolved 160 $\mu$m CIB is produced at higher redshift, with the $0.5<z\le1.0$ 
redshift bin dominating ($\sim$40\%) the budget. 

At the PEP depth, the CIB resolved into individual sources is 
mainly due to luminous and ultraluminous IR galaxies 
(LIRGs $10^{11}\le L_{IR}<10^{12}$ L$_\odot$, ULIRGs $L_{IR}\ge10^{12}$ L$_\odot$). 
The relative fraction of the two classes varies as a function of redshift, 
likely due to selection effects, with LIRGs providing 80-90\% of the CIB resolved 
at $0.5<z\le1$ and ULIRGs dominating at higher redshift 
(60-70\% at $1.0<z\le2.0$ and 100\% above).

Finally, Fig. \ref{fig:pep_cib} also reports the CIB contribution 
of 24 $\mu$m sources, as obtained through stacking (see Sect. \ref{sect:cib}). 
The relative CIB fractions emitted in the four redshift bins
change significantly at 100 $\mu$m, with respect to 
those obtained by integrating the observed counts. The redshift 
distribution of the CIB is now peaking between $0.5<z\le1.0$ and the 
relative contribution of $z>1$ increases by more than 10\%. 
On the other hand, at 160 $\mu$m only the relative fraction of the highest redshift bin 
varies by more than 5\%, while the others remain unchanged within $1-2$\%.
Also in the case of stacking, the 160 $\mu$m CIB relative redshift distribution is 
more populated at high redshift than the 100 $\mu$m one.
As expected, this indicates that
PACS galaxies with redder observed colors lie on average at higher redshift than bluer ones.


\onltab{1}{
\begin{table*}[!ht]
\centering
\begin{tabular}{r | c c | c c | c c | c c | c c | c c}
\hline
\hline
\multicolumn{1}{c|}{$S_{center}$} & \multicolumn{2}{c|}{GOODS-N 100$\mu$m} & \multicolumn{2}{c|}{LH 100$\mu$m} & \multicolumn{2}{c|}{COSMOS 100$\mu$m} & \multicolumn{2}{c|}{GOODS-N 160$\mu$m} & \multicolumn{2}{c|}{LH 160$\mu$m} & \multicolumn{2}{c}{COSMOS 160$\mu$m}\\
\hline
      2.84  &  3.19e+04   &    0.208 &      --  	 &    --    & 	    --	     &    --	&   --	     &    --	&    	--	&	-- &	 --	   &	-- \\
      3.57  &  4.67e+04   &    0.146 &      --  	 &    --    & 	    --	     &    --	&   --	     &    --	&    	--	&	-- &	 --	   &	-- \\
      4.50  &  5.57e+04   &    0.149 &      --  	 &    --    & 	    --	     &    --	&   --	     &    --	&    	--	&	-- &	 --	   &	-- \\
      5.66  &  6.37e+04   &    0.162 &      7.00e+04	 &  0.086   & 	    --	     &    --	&   1.17e+05 &    0.129 &    	--	&	-- &	 --	   &	-- \\
      7.13  &  6.90e+04   &    0.185 &      7.22e+04	 &  0.100   & 	  7.23e+04   &    0.028	&   1.42e+05 &	  0.173 &    	--	&	-- &	 --	   &	-- \\
      8.97  &  6.43e+04   &    0.234 &      6.58e+04	 &  0.128   & 	  8.10e+04   &    0.031	&   1.63e+05 &	  0.142 &    1.43e+05	&   0.086  &	 --	   &	-- \\
     11.30  &  6.29e+04   &    0.287 &      7.24e+04	 &  0.146   & 	  7.47e+04   &    0.039	&   1.74e+05 &	  0.163 &    1.67e+05	&   0.093  &	 --	   &	-- \\
     14.22  &  6.81e+04   &    0.333 &      8.13e+04	 &  0.167   & 	  6.81e+04   &    0.050	&   1.66e+05 &	  0.200 &    1.82e+05	&   0.106  &	 --	   &	-- \\
     17.91  &  1.11e+05   &    0.302 &      7.00e+04	 &  0.217   & 	  7.33e+04   &    0.057	&   1.77e+05 &	  0.237 &    1.70e+05	&   0.132  &	 2.03e+05  &	0.033\\
     22.54  &  8.13e+04   &    0.443 &      6.66e+04	 &  0.269   & 	  7.15e+04   &    0.071	&   1.97e+05 &	  0.266 &    1.53e+05	&   0.171  &	 2.07e+05  &	0.039\\
     28.38  &  1.11e+05   &    0.438 &      6.70e+04	 &  0.317   & 	  6.70e+04   &    0.087	&   1.83e+05 &	  0.334 &    1.85e+05	&   0.183  &	 1.95e+05  &	0.049\\
     35.73  &  6.63e+04   &    0.701 &      4.01e+04	 &  0.529   & 	  5.74e+04   &    0.116	&   1.88e+05 &	  0.394 &    1.66e+05	&   0.238  &	 1.80e+05  &	0.062\\
     44.98  &  6.68e+04   &    0.814 &      5.97e+04	 &  0.484   & 	  5.91e+04   &    0.135	&   1.49e+05 &	  0.527 &    1.42e+05	&   0.312  &	 1.48e+05  &	0.082\\
     56.62  &  --	  &   --     &    $\le$1.81e+04	 &  1.000   & 	  5.06e+04   &    0.175	&   1.44e+05 &	  0.641 &    1.14e+05	&   0.415  &	 1.41e+05  &	0.100\\
     71.29  &  --	  &   --     &	  $\le$2.40e+04	 &  1.000   & 	  5.64e+04   &    0.198	&   1.01e+05 &	  0.883 &    7.01e+04	&   0.658  &	 1.16e+05  &	0.135\\
     89.74  &  --	  &   --     &	    8.46e+04	 &  0.655   & 	  3.89e+04   &    0.289	&   --       &	  --    &    6.37e+04	&   0.828  &	 9.19e+04  &	0.185\\
    112.98  &  --	  &   --     &		--	 &   --	    & 	  2.51e+04   &    0.453	&   --	     &    --	&    8.19e+04	&   0.833  &	 7.70e+04  &	0.239\\
    142.23  &  --	  &   --     &		--	 &   --	    & 	  3.81e+04   &    0.412	&   --	     &    --	&       --	&       -- &	 8.38e+04  &	0.271\\
    179.06  &  --	  &   --     &		--	 &   --	    & 	    --	     &    --	&   --	     &    --	&       --	&       -- &	 6.57e+04  &	0.366\\  
\hline															
\end{tabular}
\tablefoot{Fluxes are provided in $[$mJy$]$. For each field/wavelength,
we list counts in units of $[$deg$^{-2}$ mJy$^{1.5}]$. Errors are given as relative fractions, and 
include both Poisson statistics and propagation of photometric uncertainties.}
\caption{PEP number counts, normalized to the Euclidean slope.}
\label{tab:pep_counts}
\end{table*}
}


\begin{acknowledgements}
PACS has been developed by a consortium of institutes led by MPE (Germany) and 
including UVIE (Austria); KU Leuven, CSL, IMEC (Belgium); CEA, LAM (France); 
MPIA (Germany); INAF-IFSI/OAA/OAP/OAT, LENS, SISSA (Italy); IAC (Spain). 
This development has been supported by the funding agencies BMVIT (Austria), 
ESA-PRODEX (Belgium), CEA/CNES (France), DLR (Germany), ASI/INAF (Italy), 
and CICYT/MCYT (Spain).
\end{acknowledgements}




\bibliographystyle{aa}
\bibliography{14610bib}


\Online

\begin{appendix} 
\section{PACS data}\label{sect:data}

PEP science demonstration data include the GOODS-N and Abell 2218 fields. 
In addition to these, part of the other PEP blank fields have 
already been scheduled and observed: Lockman Hole and COSMOS ($\sim$85\% of the planned depth).
Observations of all fields were carried out by adopting the intermediate speed 
(20 arcsec/s) scan-map mode. Table \ref{tab:blind_stats_match} lists the 
total exposure times already observed for these fields\footnote{See the PEP web page 
for information about the other fields in the survey: http://www.mpe.mpg.de/ir/Research/PEP/.}.

Data have been processed through the standard PACS reduction pipeline, version 2.0.1328, within the HCSS 
environment\footnote{HCSS is a joint development by the Herschel Science Ground
Segment Consortium, consisting of ESA, the NASA Herschel Science Center, and the HIFI, PACS and
SPIRE consortia} (Ott et al. \citeyear{ott2010}). Additionally, we employed 
custom procedures aimed at removing of interference patterns, tracking anomalies, 
re-centering positional offsets, and mapping.

Glitch removal is based on {\em multi-resolution median transform},
developed by \citet{starck1998} to detect faint sources in ISOCAM 
data. The signal due to real sources and glitches
show different signatures in the pixel timeline. These features are
recognized using a multi-scale transform, separating the various
frequencies of the signal. Once the glitch components are identified,
they are replaced by interpolated values in the pixel
timeline. 

PACS photometers exhibit a noise with a roughly $f^{-0.5}$ spectrum at relevant frequencies.
To remove the bulk of
the noise we apply a ``running-box'' high-pass median filter to each pixel timeline,
but mask the position of bright sources. The objects mask is produced
iteratively during the reduction by detecting sources on
the final map, and then we mask them in a double-pass mapping scheme. 
Testing shows
that this masked filtering method modifies the fluxes of point-like source by less than 5\%.

Imperfections, drifts, and errors in the pointing accuracy of the 
Herschel satellite were corrected by re-centering the data on a grid of known 
24 $\mu$m sources populating the fields. Such objects were stacked for all 
scan-legs in a given direction in a given map repetition 
(or for a subset of those, in the very large COSMOS maps).
The stacking result was 
then used to compute the average offset to be applied to 
this set of scan-legs, for a given direction, in a given map repetition. 
This procedure also implicitly corrects for small timing 
offsets between pointing information and data.
Absolute, systematic astrometric offsets turned out to be 
as high as 5 arcsec, while relative corrections between individual submaps
are approximately 1 arcsec.

Map reconstruction is done via simple image co-addition, based on
a simplified version of the ``drizzle'' method \citep{fruchter2002}.
Given the high data redundancy in the GOODS-N field,
the drop size is set to 1/8 of the input array pixel size. This corresponds to
1/5 and $\sim$1/4 of the output pixel size at 100 $\mu$m and 160 $\mu$m, 
respectively, thus reducing the correlated noise in the final map. 
Fields with lower redundancy were mapped by adopting a smaller drop size 
(1/4 of the input PACS array pixel size).
Images produced from each observation were weighted according to the
effective exposure of each pixel and co-added to produce the final maps.
The final error map was computed as the standard deviation of the weighted
mean. Owing to the nature of scan
maps, correlations exist between nearby pixels, in particular along the
scan direction. These correlations are close to uniform across the final
map, thus we derived a mean correlation correction factor which was then 
accounted for in the errors on the extracted fluxes.

PACS catalogs were extracted following two different approaches, optimized 
for the different scientific aims of the PEP project. We performed a 
blind extraction using the Starfinder PSF-fitting code
\citep{diolaiti2000a} and a guided extraction using 24 $\mu$m priors, following 
the method described in \citet{magnelli2009}. The two methods 
provide similar results: fluxes extracted in the two cases are consistent 
with each other, the prior extraction leading to slightly deeper --- 
although possibly biased --- catalogs. The number counts presented in this paper
were built on the blind catalog.
Point spread function (PSF) profiles were extracted from the final 
science maps, and turn out to have an FWHM of $\sim$7.5 and $\sim$11 arcsec
in the 100 $\mu$m and 160 $\mu$m bands, respectively. Aperture corrections 
were characterized on calibration observations of the Vesta asteroid. 
Absolute flux calibration is based on $\gamma$Dra, $\alpha$Tau, and $\alpha$CMa and 
makes use of the standard calibration file embedded in the PACS pipeline 
(version 2.0.1328). Typical absolute flux calibration errors 
are $\sim10$\% and include uncertainties on instrumental characterization, 
PSFs, and reference stars analysis.

Noise in PACS maps was measured
with random aperture extractions on residual images and compared to the 
observed S/N ratio for the detected sources.
The r.m.s. values thus obtained include both instrumental noise 
and confusion noise due to undetected sources (i.e., below the 3$\sigma$ threshold). 
This measured 1$\sigma$ noise is 1.00 mJy at 100 $\mu$m and 1.90 mJy at 160 $\mu$m for GOODS-N. Table 
\ref{tab:blind_stats_match} includes the noise properties of 
the SDP fields, as well as the Lockman Hole and COSMOS.

To quantify the reliability of extracted fluxes, the level of 
incompleteness and the fraction of spurious sources, 
Monte Carlo simulations were performed, creating 500 images and adding 20 
artificial objects onto science maps each, for a total of 10000 
sources.
Input and output fluxes are consistent with each other within a few percent.
Completeness is defined as the fraction of sources that have
been detected with a photometric accuracy of at least 50\% \citep{papovich2004}.
Spurious sources are defined as those extracted above 3$\sigma$ with an input 
flux lower than 3$\sigma(Image)$. The latter is consistent with the spurious fraction 
inferred by blindly extracting from inverted maps.
The GOODS-N blind catalog reaches 80\% completeness at $\sim$5.5 mJy and $\sim$11.0 mJy in the two 
bands, and a 30\% fraction of spurious detections at $\sim$2.5 
and $\sim$7.0 mJy, in green and red respectively (see Table \ref{tab:blind_stats_match}).  

The GOODS-N field benefits from an extensive multi-wavelength coverage. Adopting 
the \citet{grazian2006} approach, the PEP Team built a reliable 
multi-wavelength, PSF-matched database, including ACS $bviz$ 
\citep{giavalisco2004}, Flamingos $JHK$\footnote{Kindly 
reduced by Kyoungsoo Lee} and Spitzer IRAC data. Moreover, MIPS 24 $\mu$m 
\citep{magnelli2009} and \citet{barger2008} deep $U$, $Ks$, and spectroscopic 
redshifts have been added. When no spectroscopic redshifts were available, 
photometric redshifts have been derived using the EAZY code \citep{brammer2008}.

The 100 and 160 $\mu$m blind catalogs were finally linked to this 
multi-wavelength catalog through a three-band maximum likelihood procedure 
\citep{sutherland1992}, starting from the longest wavelength available 
(160 $\mu$m, PACS) and progressively matching 100 $\mu$m (PACS) and 
24 $\mu$m (Spitzer/MIPS) data. 
Table \ref{tab:blind_stats_match} summarizes the main properties of 
blind catalogs, as well as the results of the maximum-likelihood match in GOODS-N.

Confusion is a major concern in deep wide-beam observations, like far-IR
or sub-mm imaging of blank fields. PEP Herschel observations are not dispensed from 
confusion:
the high density of detected sources
hinders the extraction of fainter objects, in the so-called ``source density
confusion criterion'' (SDC, Dole et al. \citeyear{dole2003}).
Adopting the \citet{lagache2003} definition of {\em beam} (i.e., $\Omega=1.14\times\theta_{FWHM}^2$),
the source density in PACS blind catalogs is 40 beams/source at 100 $\mu$m and 
18 beams/source at 160 $\mu$m. This indicates that PEP GOODS-N is already hitting the SDC 
limit in the PACS red band, estimated to be 16.7 beams/source by \citet{dole2003}, while 
the green channel is not affected. By extrapolating integral number counts, we 
estimate that this limit will be reached at $\sim$2.0 mJy at 100 $\mu$m and $\sim$4.7 mJy 
at 160 $\mu$m.
Nevertheless, this limit is known to be rather conservative, and some authors have 
already shown that source extraction can be reliably carried out to levels as low as 
10 beam/source under favorable conditions \citep[e.g. at 24 $\mu$m,][]{magnelli2009}.

\begin{table*}[!ht]
\centering
\begin{tabular}{l c r c c c c}
\hline
\hline
Field   	& area & $t_{exp}$ & r.m.s. & $N(PACS)$ & Compl. & Spur.\\
\& band		&      & $[$h$]$   & $[$mJy$]$ & $>3\sigma$ & 80\% & 30\%\\
\hline
GOODS-N 100 	& $10^\prime\times15^\prime$ 	& 30    & 1.00	  & 291 & 5.5 & 2.5\\
GOODS-N 160 	& $10^\prime\times15^\prime$ 	& 30    & 1.90	  & 317 & 11.0 & 7.0\\
100+160		& 		--		& --	& --  	  & 201 & --	& --  	  \\
100+multi	& 		--		& --	& --  	  & 254 & --	& --  	  \\
160+multi	& 		--		& --	& --  	  & 274 & --	& --  	  \\
100+160+multi	& 		--		& --	& --  	  & 187 & --	& --  	  \\
100+multi+zspec & 		--		& --	& --  	  & 162 & --	& --  	  \\
160+multi+zspec & 		--		& --	& --  	  & 169 & --	& --  	  \\
100+160+multi+zspec  &		--		& --	& --  	  & 125 & --	& --  	  \\
\hline
A2218 100	& $4^\prime\times4^\prime$	& 13    & 0.84	  &  98 & 5.4 & 3.0 \\
A2218 160	& $4^\prime\times4^\prime$	& 13    & 1.59	  &  94 & 11.8 & 7.6 \\
LH\tablefootmark{a} 100		& $24^\prime\times24^\prime$	& 35    & $\sim$1.3	  &  $\sim$780   & $\sim$7.0 & $\sim$4.0 \\
LH\tablefootmark{a} 160		& $24^\prime\times24^\prime$	& 35    & $\sim$2.7	  &  $\sim$700   & $\sim$14.5 & $\sim$9.5 \\
COSMOS\tablefootmark{b} 100      & $85^\prime\times85^\prime$    & 182   & $\sim$2.0	  &  $\sim$5750   & $\sim$9.5 & $\sim$6.3\\
COSMOS\tablefootmark{b} 160      & $85^\prime\times85^\prime$    & 182   & $\sim$4.0	  &  $\sim$4900   & $\sim$20.5 & $\sim$12.0\\
\hline
\end{tabular}
\tablefoot{
\tablefoottext{a}{The PEP Lockman Hole field is the ROSAT-HRI and XMM field.}
\tablefoottext{b}{COSMOS will reach 213 hours of integration at full depth.}
}
\caption{PEP fields: total exposure times, noise properties, flux levels for 
80\% completeness and 30\% spurious fraction,
statistics of \textit{blind} catalogs, and results of maximum-likelihood 
match to the multi-wavelength ancillary catalogs (labeled ``multi'', GOODS-N only).}
\label{tab:blind_stats_match}
\end{table*}

\section{Authors affiliations}\label{sect:affiliations}

\begin{enumerate}[label=$^{\arabic{*}}$]
\item Max-Planck-Institut f\"{u}r Extraterrestrische Physik (MPE),
Postfach 1312, 85741 Garching, Germany.
\item Herschel Science Centre, ESAC, Villanueva de la Ca\~nada, 28691 Madrid, Spain.
\item Laboratoire AIM, CEA/DSM-CNRS-Universit{\'e} Paris Diderot, IRFU/Service d'Astrophysique, 
B\^at.709, CEA-Saclay, 91191 Gif-sur-Yvette Cedex, France.
\item ESO, Karl-Schwarzschild-Str. 2, D-85748 Garching, Germany.
\item INAF - Osservatorio Astronomico di Trieste, via Tiepolo 11, 34143 
Trieste, Italy.
\item Instituto de Astrof{\'\i}sica de Canarias, 38205 La Laguna, Spain. 
\item Departamento de Astrof{\'\i}sica, Universidad de La Laguna, Spain.
\item Dipartimento di Astronomia, Universit{\`a} di Bologna, Via Ranzani 1,
40127 Bologna, Italy.
\item INAF-Osservatorio Astronomico di Bologna, via Ranzani 1, I-40127 Bologna, Italy.
\item INAF - Osservatorio Astronomico di Roma, via di Frascati 33, 00040 Monte Porzio Catone, Italy.
\item Dipartimento di Astronomia, Universit{\`a} di Padova, Vicolo dell'Osservatorio 3, 
35122 Padova, Italy.
\end{enumerate}

\end{appendix}


\end{document}